\documentclass[11pt]{article}
\usepackage[article,hylinks]{subhas}
\RequirePackage[letterpaper,hmargin=1.25in,vmargin=1.25in]{geometry}
 
\newcommand{\PPSZ}{\lang{PPSZ}}
\newcommand{\PPZ}{\lang{PPZ}}
\newcommand{\DEL}{\lang{DEL}}
\newcommand{\DELPPZ}{\lang{DEL}\text{-}\lang{PPZ}}
\newcommand{\TWSAT}{\lang{2SAT}}

\newcommand{\CNF}{\lang{CNF}}

\newcommand{\UKSAT}{\lang{Unique}\text{-}k\text{-}\SAT}
\newcommand{\UTSAT}{\lang{Unique}\text{-}3\text{-}\SAT}
\begin{document}

\title{A Randomized Algorithm for 3-SAT}
\myauthor

\maketitle              
\begin{abstract}
In this work we propose and analyze a simple randomized algorithm to find a satisfiable assignment for a Boolean formula in conjunctive normal form ($\CNF$) having at most $3$ literals in every clause. Given a $k$-$\CNF$ formula $\phi$ on $n$ variables, and $\alpha \in \set{0,1}^n$ that satisfies $\phi$, a clause of $\phi$ is critical if exactly one literal of that clause is satisfied under assignment $\alpha$. Paturi et. al. (Chicago Journal of Theoretical Computer Science 1999) proposed a simple randomized algorithm ($\PPZ$) for $k$-$\SAT$ for which success probability increases with the number of critical clauses (with respect to a fixed satisfiable solution of the input formula). Here, we first describe another simple randomized algorithm $\DEL$ which performs better if the number of critical clauses are less (with respect to a fixed satisfiable solution of the input formula). Subsequently, we combine these two simple algorithms such that the success probability of the combined algorithm is maximum of the success probabilities of $\PPZ$ and $\DEL$ on every input instance. We show that when the average number of clauses per variable that appear as unique true literal in one or more critical clauses in $\phi$ is between $1$ and $2/(3 \cdot \log{(3/2)})$, combined algorithm performs better than the $\PPZ$ algorithm.
\end{abstract}
\section{Introduction}
The problem of finding a satisfiable assignment ($\SAT$) for a propositional formula in conjunctive normal form ($\CNF$) is notably the most important problem in the theory of computation. The decision problem for $\CNF$-$\SAT$ was one of the first problems shown to be $\NP$-complete\cite{Cook71,Levin73}.  $\CNF$-$\SAT$ is widely believed to require deterministic algorithm of exponential time complexity. A syntactically restricted version of general $\CNF$-$\SAT$ is $k$-$\SAT$, where each clause of a given $\CNF$ formula contains at most $k$ literals, for some constant $k$. $k$-$\SAT$ remains $\NP$ complete for $k \geq 3$ (while $2$-$\SAT$ is solvable in polynomial time \cite{APT79}). This restriction on the number of literals per clause seem to be of help, and existing algorithms have ${\mathcal O}\left(2^{\epsilon_k n}\right)$ time complexity for some constant $0 < \epsilon_k < 1$ dependent on $k$. Several work exists on faster algorithms for $k$-$\SAT$ (cf. \cite{Dan81}, \cite{MS85}, \cite{Sch02}, \cite{DGHKKPRS02}, \cite{Paturi05}).\par
The objectives of working on $k$-$\SAT$ algorithms are several. Primary of them is to obtain algorithms having provable bounds on the running time that is significantly better than trivial search algorithm (which is $\poly\left(n\right)2^{n}$ for formula having $n$ variables) and works for larger set of $k$-$\CNF$. Second objective is to understand instances that are significantly hard or easy while useful (i.e. they appear in practical problems).\par
In following we mention all bounds by suppressing the polynomial factors. Monien and Speckenmeyer \cite{MS85} described first such non-trivial algorithm with running time ${\mathcal O}\left(2^{\left(1-\epsilon_k\right)n}\right)$, with $\epsilon_k > 0$ for all $k$, and in specific it is ${\mathcal O}\left(1.618^n\right)$ for $k=3$. Faster algorithm for $3$-$\CNF$ satisfiability is due to Kullmann \cite{Kul99}, with running time ${\mathcal O}\left(1.505^n\right)$ for $k=3$. Both of these algorithms are deterministic. Paturi et al. \cite{PPZ99} proposed a simple \emph{randomized algorithm} for $k$-$\SAT$. Though it is not faster than other known algorithms for $k = 3$, it has better performance for larger values of $k$. This algorithm was improved in \cite{PPSZ98,Paturi05} with a randomized variant of the Davis-Putnam procedure \cite{DLL62} with limited resolution. Sch\"{o}ning's random walk algorithm \cite{Uwe99,Sch02} is better than \cite{Paturi05} for $k = 3$, but is worse for $k \geq 4$. Sch\"{o}ning's random walk algorithm \cite{Sch02} has bound of ${\mathcal O}\left(\left(2-{2}/{\left(k+\epsilon\right)}\right)^n\right)$ for some $\epsilon > 0$. Further improvements of his algorithm were found by Hofmeister et al. \cite{HSSW07} for $k=3$. Randomized algorithm of \cite{Paturi05} has expected running time ${\mathcal O}\left(1.362^n\right)$ for $k=3$. \par
Better randomized algorithm is due to Iwama and Tamaki \cite{IT04}, having expected running time ${\mathcal O}\left(1.3238^n\right)$ for $k=3$, which is a combination of the Sch\"{o}ning's random walk algorithm \cite{Uwe99,Sch02} and the algorithm of Paturi et al. \cite{PPSZ98} (this bound improves to ${\mathcal O}\left(1.32266^n\right)$ using modified analysis in \cite{Paturi05}). Iwama and Tamaki's algorithm \cite{IT04} has been improved by Rolf \cite{Rolf06} recently to best known randomized bound of ${\mathcal O}\left(1.32216^n\right)$ for $3$-$\SAT$. \par
Sch\"{o}ning's algorithm was derandomized in \cite{DGHKKPRS02} to the currently best known bound of ${\mathcal O}\left(1.481^n\right)$ for $k=3$ and to a bound of ${\mathcal O}\left(\left(2-{2}/{\left(\left(k+1\right)+\epsilon\right)}\right)^n\right)$ for $k>3$, using limited local search and covering codes. This was improved for  $k=3$ in \cite{BK04} to a deterministic bound of ${\mathcal O}\left(1.473^n\right)$. Randomized algorithm of \cite{PPSZ98} was derandomized in \cite{Rol05} for $\UKSAT$ (i.e. $k$-$\CNF$ formulas having only one solution) using techniques of limited independence, i.e. by constructing a small bias probability space to choose samples for original algorithm of \cite{PPSZ98} yielding deterministic running time ${\mathcal O}\left(1.3071^n\right)$ for $\UTSAT$.
In this work we present and analyze a randomized algorithm for finding a satisfiable assignment for a Boolean formula in $\CNF$ having at most $3$ literals in every clause. We consider the $k$-$\SAT$ algorithm of Paturi et al. \cite{PPZ99} for $k = 3$ and combine it with another randomized algorithm that we describe here, such that the success probability of the combined algorithm is maximum of the success probabilities of these two algorithms on every input instance.\par
Before we proceed further let us introduce some notations. A formula $\phi$ in $n$-variables is defined over a set $\set{x_1,\ldots, x_n}$. \emph{Literals} are variable $x$ or negated variable $\neg x$. \emph{Clauses} are disjunctions of literals, and we assume that a clause do not contain both, a literal and its negation. A Boolean formula $\phi = \wedge_{i = 1}^{m}{C_i}$ is a $k$-$\CNF$ if each clause $C_i$ is a disjunction of at most $k$ literals. Variables are assigned truth values $1$ (\true) or $0$ (\false). An assignment to variables $\set{x_1,\ldots, x_n}$ is an element $\alpha \in \set{0,1}^n$. For $S \subseteq \set{0,1}^n$ and $\alpha \in S$, $\alpha$ is an \emph{isolated} point of $S$ in direction $i$ if flipping $i$th bit of $\alpha$ produces an element that is not in $S$. We will call $\alpha \in S$, $j$--isolated in $S$ if there are exactly $(n-j)$ neighbors of $\alpha$ in $S$. An $n$-isolated point in $S \subseteq \set{0,1}^n$ will be called isolated.\par
\begin{algorithm}
\dontprintsemicolon
\textbf{Algorithm} $\PPZ\left(\phi\right)$ \text{ Input: } $3$-$\CNF$ $\phi = \wedge_{i = 1}^{m}{C_i}$ \text{ on variables } $\set{x_1, \ldots, x_n}$\\

Pick a permutation $\pi$ of the set $\set{1,\ldots,n}$ uniformly at random.\;
\For{$i = 1, \ldots, n$}
{
	\eIf{there is an unit clause corresponding to the variable $x_{\pi(i)}$}
	{
		Set $x_{\pi(i)}$ so that corresponding unit clause is satisfied, let $b$ be the assignment.\;
	}
	{
		Set $x_{\pi(i)}$ to $\true$ or $\false$ uniformly at random, let $b$ be the assignment.\;
	}
	$\phi := \phi[x_{\pi(i)} \leftarrow b]$, $\alpha_i := b$.\;
}
\eIf{$\alpha$ is a satisfying assignment}
{
	\Return{$\alpha$}.\;
}{
	\Return{``Unsatisfiable''}.\;
}

\caption{One iteration of procedure $\PPZ\left(\phi\right)$}
\label{PPZ}
\end{algorithm}
Given a $k$-$\CNF$ formula $\phi$ on $n$ variables $\set{x_1, \ldots, x_n}$, single iteration of Paturi et al.'s randomized algorithm \cite{PPZ99} (see Algorithm-\ref{PPZ}) works by selecting a random permutation of variables $\pi \in S_n$, and then assigning truth values uniformly at random in $\set{0,1}$ to each variable $x_{\pi(i)}$ for $i = 1, \ldots, n$. However, before assigning a random truth value, algorithm checks if there is an unsatisfied unit clause (i.e., a clause having only one literal) corresponding to variable $x_{\pi(i)}$, and if there is one, it forces the value of $x_{\pi(i)}$ such that the corresponding unit clause gets satisfied. We will call this algorithm $\func{PPZ}$. Let $S \subseteq \set{0,1}^n$ be the set of all satisfying assignments of $\phi$.\par
Crucial observation made in \cite{PPZ99} is that if $\alpha$ is an isolated point of $S$ in some direction $i$, then there exists a clause in which exactly one literal is satisfied under assignment $\alpha$ -- and that literal corresponds to the variable $x_i$ (such a clause will be called \emph{critical} for variable $x_i$ under solution $\alpha$). Given formula $\phi$ let $\alpha \in S$ be a fixed satisfying assignment in the set of all satisfying assignments of $\phi$. Now observe that after selecting a random permutation of variables $\pi$, probability that $\func{PPZ}(\phi)$ outputs assignment $\alpha$ depends on number of variables that are not forced. On the other hand variables that are forced correspond to at least one critical clause. Thus $\pr[\PPZ(\phi) = \alpha | \pi]$ improves if there are more critical clauses. With clever analysis it was shown in \cite{PPZ99} that the success probability that one iteration of $\PPZ$ finds a satisfying assignment of $\phi$ is at least $2^{-n(1-1/k)}$ -- which is at least $2^{-2n/3}$ for $3$-$\CNF$. Finally we note that $\PPZ$ makes one-sided error - if input formula $\phi$ is unsatisfiable then algorithm will always say so, but on satisfiable instances it may make error.\par 
Let us consider another very simple randomized algorithm for $3$-$\CNF$. We will call this algorithm $\DEL$ (see Algorithm-\ref{DEL}). In a single iteration of this algorithm we first delete one literal from each clause having three literals independently uniformly at random (a clause having less than three literals is ignored in this step) and obtain a new formula. Since input formula $\phi$ is a $3$-$\CNF$, we obtain a new formula $\phi'$ in $2$-$\CNF$ for which there is a known linear time deterministic algorithm \cite{APT79} (we will call this algorithm $\TWSAT$). After running algorithm $\TWSAT(\phi')$ if we find a satisfying assignment then we output that (after extending it to the rest of the variables (if any) - which can be assigned any truth value).\par
\begin{algorithm}
\dontprintsemicolon
\textbf{Algorithm} $\DEL\left(\phi\right)$ \text{ Input: } $3$-$\CNF$ $\phi = \wedge_{i = 1}^{m}{C_i}$ \text{ on variables } $\set{x_1, \ldots, x_n}$\\

\For(\tcc*[f]{ignore clause with less than $3$ literals}){Each clause $C$ having $3$ literals}{
Select one literal uniformly at random and delete it.\;
}
Let $\phi'$ be the obtained $2$-$\CNF$.\;
\eIf{$\TWSAT(\phi')$ returns a satisfiable assignment $\alpha$}
{
		 \Return{$\alpha$}.\;
}{
	\Return{``Unsatisfiable''}.\;
}
\caption{One iteration of procedure $\DEL\left(\phi\right)$}
\label{DEL}
\end{algorithm}
Again, let $\alpha \in S$ be a fixed solution in the set of all solutions of the input formula $\phi$. Let $C(\alpha)$ be a critical clause of $\phi$ for variable $x$ under solution $\alpha$. Now observe that in the process of deletion if we delete the literal corresponding to variable $x$ from $C(\alpha)$ then in the first step of the algorithm $\DEL(\phi)$ we may produce a formula $\phi'$ having no satisfying assignment (e.g. when $\alpha$ is the unique solution of formula $\phi$, or if we make this error in a critical clause with respect to an isolated solution). Probability that this event does not happen is $2/3$ for $C(\alpha)$ - as a clause can not be critical for more than one variable, and every clause have $3$ literals (other clauses with less than three literals were not considered in the deletion step). Now observe that only the deletion step of the algorithm $\DEL$ makes randomized choices, while executing the algorithm $\TWSAT$ on $\phi'$ is deterministic. Hence, if the deletion step of the algorithm makes no error (i.e. it does not remove solutions) then algorithm $\TWSAT$ on $\phi'$ will always find a satisfying assignment whenever input formula $\phi$ is satisfiable. Now assume there are $c(\alpha)$ number of critical clauses of $\phi$ under solution $\alpha$. Then we have the probability that $\DEL(\phi)$ returns a satisfying assignment with respect to an $\alpha \in S$ is $(2/3)^{c(\alpha)}$. In general $c(\alpha)$ can be polynomial in $n$, thus $\DEL$ performs well only when all satisfiable solutions of $\phi$ have less number of critical clauses. Let us note that like $\PPZ$ algorithm, $\DEL$ also makes one-sided error - if input formula $\phi$ is unsatisfiable then algorithm will always say so, but on satisfiable instances it may make error. This can be seen from the following: assume that the input formula $\phi$ is unsatisfiable but $\TWSAT(\phi')$ returns with a satisfiable assignment - but $\phi'$ is obtained from $\phi$ by deleting one literal from each clause of size three, and hence the assignment that satisfies $\phi'$ also satisfies $\phi$ - a contradiction.\par
While success probability of $\DEL$ decreases with increasing number of critical clauses with respect to a fixed satisfiable solution $\alpha$ -- success probability of $\PPZ$ increases. This fact suggests that a combination of these two algorithms can perform better. In order to motivate this further consider the worst case of $\PPZ$ algorithm \cite{PPZ99} on $3$-$\CNF$. One such example is $\phi = \wedge_{i = 0}^{m-1}{(x_{3i+1} \oplus x_{3i+2} \oplus x_{3i+3})}$ where $n = 3m$. Any solution $\alpha$ of $\phi$ has $n$ critical clauses with respect to $\alpha$, e.g. $\{(x_{3i+1} + \bar{x}_{3i+2} + \bar{x}_{3i+3})$, $(\bar{x}_{3i+1} + {x}_{3i+2} + \bar{x}_{3i+3})$, $(\bar{x}_{3i+1} + \bar{x}_{3i+2} + {x}_{3i+3})\}_{i = 0}^{m-1}$, and success probability of $\PPZ$ on $\phi$ is $2^{-2n/3} \geq (1.5875)^{-n}$. On the other hand success probability of $\DEL$ on this instance is $(2/3)^{n} = (1.5)^{-n}$, and this is more than the success probability of $\PPZ$. Our objective in this work is to combine these two algorithms such that the success probability of the combined algorithm is maximum of the success probability of $\DEL$ and $\PPZ$ on every input instance.
\paragraph{Organization.} Rest of the paper is organized as follows. In section-\ref{SEC2} we describe the algorithm $\DELPPZ$ - which is a combination of algorithm $\PPZ$ and algorithm $\DEL$ described before. Subsequently, in section-\ref{SEC3} we analyze this combined algorithm. Finally in section-\ref{SEC4} we conclude the paper.

\section{Combined algorithm}
\label{SEC2}
In this section we describe the algorithm $\DELPPZ$ (see Algorithm-\ref{DELPPZ}) -- which is a combination of the algorithm $\PPZ$ and algorithm $\DEL$ described above. Algorithm-\ref{DELPPZ} describes one iteration, and in order to increase the success probability as a standard technique the algorithm needs to be executed several times. We will discuss about it at the end of this section. Like $\PPZ$, one iteration of $\DELPPZ$ algorithm works by first selecting a random permutation of variables $\pi \in S_n$. Then for $i = 1, \ldots, n$ the algorithm either execute steps that are similar to $\DEL(\phi)$ and, if unsuccessful in finding a satisfying assignment, it execute steps that are similar to $\PPZ$.\par
\begin{algorithm}
\dontprintsemicolon
\textbf{Algorithm} $\DELPPZ\left(\phi\right)$ \text{ Input: } $3-\CNF$ $\phi = \wedge_{i = 1}^{m}{C_i}$ \text{ on variables } $\set{x_1, \ldots, x_n}$\\

Pick a permutation $\pi$ of the set $\set{1,\ldots,n}$ uniformly at random.\;
$\alpha := 0^n$\;
\For{$i = 1, \ldots, n$}
{
	
	\For(\tcc*[f]{ignore clause with less than $3$ literals}){Each clause $C$ having $3$ literals}{
	Select one literal uniformly at random and delete it.\;
	}
	Let $\phi'$ be the obtained $2$-$\CNF$.\;
	\eIf{$\TWSAT(\phi')$ returns a satisfiable assignment $\beta$}
	{
		 $(*)$ \Return{$\beta$}.\;
	}
	{
		\eIf{there is an unit clause corresponding to the variable $x_{\pi(i)}$}
		{
			Set $x_{\pi(i)}$ so that corresponding unit clause is satisfied, let $b$ be the assignment.\;
		}
		{
			Set $x_{\pi(i)}$ to $\true$ or $\false$ uniformly at random, let $b$ be the assignment.\;
		}
	}
	$\phi := \phi[x_{\pi(i)} \leftarrow b]$, $\alpha_i := b$.\;
}
\eIf{$\alpha$ is a satisfying assignment}
{
	$(**)$ \Return{$\alpha$}.\;
}{
	\Return{``Unsatisfiable''}.\;
}

\caption{One iteration of procedure $\DELPPZ\left(\phi\right)$}
\label{DELPPZ}
\end{algorithm}
In other words, for each $i = 1, \ldots, n$ the algorithm works on the current formula $\phi$ (like $\PPZ$, input formula $\phi$ is modified in every execution of the for loop as we assign truth value to variable $x_{\pi(i)}$ in $i$th execution) and first delete one literal from each clause of $\phi$ having three literals independently uniformly at random (a clause having less than three literals is ignored in this step) and obtain a new formula $\phi'$. Since input formula $\phi$ is a $3$-$\CNF$, we obtain a new formula $\phi'$ in $2$-$\CNF$. After running algorithm $\TWSAT(\phi')$ if we find a satisfying assignment then we output that (after extending it to the rest of the variables -- which can be assigned any truth value), or else we again consider the current formula $\phi$ and assign truth values in $\set{0,1}$ to variable $x_{\pi(i)}$. This is done as follows: we first check if there is an unsatisfied unit clause corresponding to variable $x_{\pi(i)}$ and force the value of $x_{\pi(i)}$ such that the corresponding unit clause gets satisfied, otherwise we assign truth values in $\set{0,1}$ to $x_{\pi(i)}$ uniformly at random.\par
After this, the current formula $\phi$ is modified as $\phi := \phi[x_{\pi(i)} \leftarrow b]$. Where, by $\phi := \phi[x_{\pi(i)} \leftarrow b]$ we denote that variable $x_{\pi(i)}$ is assigned $b \in \set{0,1}$, and formula $\phi$ is modified by treating each clause $C$ of $\phi$ as follows: $(i)$ if $C$ is satisfied with this assignment then delete $C$, otherwise $(ii)$ replace clause $C$ by clause $C'$ obtained by deleting any literals of $C$ that are set to $0$ by this assignment. Hence, $\DEL(\phi)$ works on a new instance of formula in each execution of the for loop.\par
In every execution there are two places from where the algorithm could exit and return a satisfying assignment. When $\TWSAT(\phi')$ returns a satisfying assignment $\beta$ for some $i = 1, \ldots, n$(marked as $(*)$, and we shall call it return by $\DEL$) or at the end (marked as $(**)$, which we shall call as return by $\PPZ$).\par
It is not hard to see that the algorithm $\DELPPZ$ never returns an assignment if the input formula is unsatisfiable. As stated earlier, both $\PPZ$ and $\DEL$ has one-sided error and similar argument holds for $\DELPPZ$ as well. Thus the problem of interest would be to bound the probability that the algorithm answers ``unsatisfiable'' when the input formula $\phi$ is satisfiable. If $\tau(\phi)$ is the success probability of the algorithm $\DELPPZ$ on input $\phi$, and if we execute the algorithm $\omega$ number of times, then for a satisfiable formula $\phi$ the error probability is equal to $(1-\tau(\phi))^{\omega} \leq e^{-(\omega \cdot \tau(\phi))}$. This will be at most $e^{-n}$ if we choose  $\omega \geq n/\tau(\phi)$. In following section we shall estimate $\tau(\phi)$ and subsequently choose the value of $\omega$. 
\section{Analysis of the combined algorithm}
\label{SEC3}
In this section we analyze the algorithm $\DELPPZ$. Let $\phi = \wedge_{i = 1}^{m}{C_i}$ be the input $3-\CNF$ formula defined on $n$ variables $\set{x_1, \ldots, x_n}$. Let $S \subseteq \set{0,1}^n$ be the set of satisfying assignments of $\phi$, $\alpha \in S$, and let $\pi$ be any permutation in $S_n$. \par
Observe that in the main loop for each $i = 1, \ldots, n$, the algorithm can return by $\DEL$ (marked as $(*)$) for any $i$. When the algorithm returns by $\DEL$ in the $i$th execution of the for loop, we estimate the success probability of obtaining any satisfying assignment in that execution of the for loop with respect to a $\alpha \in S$, for a fixed $\pi \in S_n$. Let us denote the $i$th such event by $A_{i}(\alpha)$ for $i = 1, \ldots, n$ to indicate that $i$th execution returns by $\DEL$ with some satisfying assignment. To indicate that $\pi \in S_n$ is fixed we use the shorthand notation $\pr[A | \pi]$ to denote $\pr[A | \text{When } \pi \text{ is fixed}]$, for some event $A$. Also, for any event $A$ let $\overline{A}$ denote the complement of event $A$.\par
Similarly, let the event $B$ denote that the algorithm returns by $\PPZ$ at the end of the for loop (marked as $(**)$) and satisfying assignment returned is $\alpha$, again for a fixed $\pi \in S_n$. Let us denote by $\DELPPZ(\phi, \alpha)$ the event that with respect to some $\alpha \in S$, algorithm $\DELPPZ$ returns with a successful satisfying assignment - either by $\DEL$ or by $\PPZ$. Now observe that the algorithm either returns by $\DEL$ in any one of the execution of the for loop for $i = 1 \ldots, n$, or it returns by $\PPZ$ at the end of the for loop, hence, $\pr[(\cup_{i=1}^{n}A_i(\alpha)) \cap B | \pi] = 0$. With this we have:
\begin{align}
&\pr[\DELPPZ(\phi, \alpha)| \pi] = \pr[\bigcup\limits_{i=1}^{n}A_i(\alpha) \vee B | \pi] =\notag\\
&\left(\sum\limits_{i = 1}^{n}{\pr[A_i(\alpha) | \bigwedge\limits_{j = 1}^{i-1}\overline{A_j(\alpha)} \wedge \pi] \cdot \pr[\bigwedge\limits_{j = 1}^{i-1}\overline{A_j(\alpha)} | \pi]}\right) +\notag\\
& \pr[B | \bigwedge\limits_{i = 1}^{n}\overline{A_i(\alpha)} \wedge \pi] \cdot \pr[\bigwedge\limits_{i = 1}^{n}\overline{A_i(\alpha)} | \pi]\label{EQ1}
\end{align}
Recall, if the deletion step of the algorithm makes no error then algorithm $\TWSAT$ on $\phi'$ will always find a satisfying assignment. On the other hand in the process of deletion if we delete any unique true literal corresponding to a critical clause with respect to satisfying assignment $\alpha$ we may produce a formula $\phi'$ which will not have any satisfying assignment, and we will make error.\par
Let $c_{\pi}^{i-1}(\alpha)$ be the number of critical clauses of the resulting formula in the $i$th step with respect to assignment $\alpha$ on which the deletion step of $\DEL$ and subsequently $\TWSAT$ is executed. In specific $c_{\pi}^{0}(\alpha)$ denotes the number of critical clauses of the input formula $\phi$. Since $c_{\pi}^{i-1}(\alpha)$ is the number of critical clauses of the resulting formula used in the $i$th step with respect to assignment $\alpha$ then success probability of returning by $\DEL$ in that step i.e. $\pr[A_i(\alpha) | \wedge_{j = 1}^{i-1}\overline{A_j(\alpha)} \wedge \pi]$ is $(2/3)^{c_{\pi}^{i-1}(\alpha)}$. Now for collection of events $\overline{A_1(\alpha)}, \ldots, \overline{A_n(\alpha)}$ it holds that, for $r = 1, \ldots, n$,
\begin{align}
&\pr[\bigwedge\limits_{j = 1}^{r}\overline{A_i(\alpha)} | \pi] = \pr[\overline{A_1(\alpha)} | \pi] \cdot \pr[\overline{A_2(\alpha)} | \overline{A_1(\alpha)} \wedge \pi] \cdot \ldots\notag\\
& \cdot \pr[\overline{A_{r}(\alpha)} | \bigcap\limits_{j = 1}^{r-1}{\overline{A_j(\alpha)}} \wedge \pi]\notag
\end{align}
Observe that if the algorithm fails to return by $\DEL$ in the $(r-1)$th execution of the for loop, then given there were $c_{\pi}^{r-2}(\alpha)$ many critical clauses in the beginning of the $(r-1)$th execution, there will be $c_{\pi}^{r-1}(\alpha)$ many critical clauses after $\PPZ$ part of the algorithm executes. Hence, for $r = 1, \ldots, n,$ given all $(r-1)$ trial of return by $\DEL$ has failed we have:
\begin{align}
\pr[\overline{A_{r}(\alpha)} | \bigcap\limits_{j = 1}^{r-1}{\overline{A_j(\alpha)}} \wedge \pi] &= \left(1-\left(\frac{2}{3}\right)^{c_{\pi}^{r-1}(\alpha)}\right), \text{ for } r = 1, \ldots, n.\notag
\end{align}
Hence, 
\begin{align}
\pr[\bigwedge\limits_{j = 1}^{r}\overline{A_i(\alpha)} | \pi] &=  \prod\limits_{j = 1}^{r}{\left(1-\left(\frac{2}{3}\right)^{c_{\pi}^{r-1}(\alpha)}\right)}, \text{ for } r = 1, \ldots, n.\notag
\end{align}
And we have,
\begin{align}
&\sum\limits_{i = 1}^{n}{\pr[A_i(\alpha) | \bigwedge\limits_{j = 1}^{i-1}\overline{A_j(\alpha)} \wedge \pi] \cdot \pr[\bigwedge\limits_{j = 1}^{i-1}\overline{A_j(\alpha)} | \pi]} =\notag\\
&\sum\limits_{i = 1}^{n}{\left(\frac{2}{3}\right)^{c_{\pi}^{i-1}(\alpha)} \cdot \prod\limits_{j = 1}^{i-1}{\left(1-\left(\frac{2}{3}\right)^{c_{\pi}^{j-1}(\alpha)}\right)}}\label{EQ2}
\end{align}
Let $d_{\pi}(\alpha)$ be the number of variables that are \emph{not} forced by $\PPZ$. Then we have:
\begin{align}
\pr[B | \bigwedge\limits_{i = 1}^{n}\overline{A_i(\alpha)} \wedge \pi] \cdot \pr[\bigwedge\limits_{i = 1}^{n}\overline{A_i(\alpha)} | \pi] = 2^{-d_{\pi}(\alpha)} \cdot \prod\limits_{i = 1}^{n}{\left(1-\left(\frac{2}{3}\right)^{c_{\pi}^{i-1}(\alpha)}\right)}\label{EQ3}
\end{align}
Using Eq. (\ref{EQ2}) and Eq. (\ref{EQ3}) with Eq. (\ref{EQ1}) it is easy to see now that,
\begin{align}
\pr[\DELPPZ(\phi,\alpha) | \pi] &= \sum\limits_{i = 1}^{n}{\left(\left(\frac{2}{3}\right)^{c_{\pi}^{i-1}(\alpha)} \cdot \prod\limits_{j = 1}^{i-1}{\left(1-\left(\frac{2}{3}\right)^{c_{\pi}^{j-1}(\alpha)}\right)}\right)} +\notag\\
& \left(2^{-d_{\pi}(\alpha)} \cdot \prod\limits_{i = 1}^{n}{\left(1-\left(\frac{2}{3}\right)^{c_{\pi}^{i-1}(\alpha)}\right)}\right)\label{EQ4}
\end{align}
Let $\av_{\pi}[X]$ denote the expectation of random variable $X$ taken over all random permutation $\pi \in S_n$. Now it is easy to see that using Eq. (\ref{EQ4}), and summing over the set $S$ of all satisfying solutions of $\phi$, we have using linearity of expectation:
\begin{align}
\tau(\phi) &= \pr[\DELPPZ(\phi) \text{ outputs some satisfying assignment}]\notag\\
&= \sum\limits_{\alpha \in S}{\av_{\pi}\left[\sum\limits_{i = 1}^{n}{\left(\left(\frac{2}{3}\right)^{c_{\pi}^{i-1}(\alpha)} \cdot \prod\limits_{j = 1}^{i-1}{\left(1-\left(\frac{2}{3}\right)^{c_{\pi}^{j-1}(\alpha)}\right)}\right)}\right]} + \notag\\
&\sum\limits_{\alpha \in S}{\av_{\pi}\left[2^{-d_{\pi}(\alpha)} \cdot \prod\limits_{i = 1}^{n}{\left(1-\left(\frac{2}{3}\right)^{c_{\pi}^{i-1}(\alpha)}\right)}\right]}\notag\\
&\geq \sum\limits_{\alpha \in S}{\left[\sum\limits_{i = 1}^{n}{\left(\left(\frac{2}{3}\right)^{\av_{\pi}[c_{\pi}^{i-1}(\alpha)]} \cdot \prod\limits_{j = 1}^{i-1}{\left(1-\left(\frac{2}{3}\right)^{\av_{\pi}[c_{\pi}^{j-1}(\alpha)]}\right)}\right)}\right]} + \notag\\
&\sum\limits_{\alpha \in S}{\left[2^{-\av_{\pi}[d_{\pi}(\alpha)]} \cdot \prod\limits_{i = 1}^{n}{\left(1-\left(\frac{2}{3}\right)^{\av_{\pi}[c_{\pi}^{i-1}(\alpha)]}\right)}\right]}\label{EQ5}
\end{align}
Where last inequality (Eq. (\ref{EQ5})) follows from Jensen's inequality (cf. \cite{Feller71})- which states that for a random variable $X = (c_{\pi}^{0}(\alpha), c_{\pi}^{1}(\alpha), \ldots, c_{\pi}^{n-1}(\alpha), d_{\pi}(\alpha))$ and any convex function $f$, $\av[f(X)] \geq f(\av[X])$. Now observe that $c_{\pi}^{0}(\alpha)$, $c_{\pi}^{1}(\alpha)$, $\ldots$,$ c_{\pi}^{n-1}(\alpha)$ is a non-increasing sequence of integers, i.e. $c_{\pi}^{0}(\alpha) \geq c_{\pi}^{1}(\alpha) \geq \ldots \geq c_{\pi}^{n-1}(\alpha)$, because in every execution whenever a variable is forced by $\PPZ$ a collection of critical clause gets satisfied and are removed from $\phi$. Hence, we can simplify Eq. (\ref{EQ5}) as follows using the fact that $\av_{\pi}[c_{\pi}^{0}(\alpha)] = c_{\pi}^{0}(\alpha)$, when $c_{\pi}^{0}(\alpha) \neq 0$. On the other hand when $c_{\pi}^{0}(\alpha) = 0$, it follows that $\tau(\phi) = 1$ by taking $c_{\pi}^{i-1}(\alpha) = 0$ for all $i = 1, \ldots, n$ in Eq. (\ref{EQ4}): 
\begin{align}
\tau(\phi) &\geq \sum\limits_{\alpha \in S}{\left[\left(\frac{2}{3}\right)^{c_{\pi}^{0}(\alpha)} \cdot \sum\limits_{i = 1}^{n}{\prod\limits_{j = 1}^{i-1}{\left(1-\left(\frac{2}{3}\right)^{\av_{\pi}[c_{\pi}^{j-1}(\alpha)]}\right)}}\right]} + \notag\\
&\sum\limits_{\alpha \in S}{\left[2^{-\av_{\pi}[d_{\pi}(\alpha)]} \cdot \prod\limits_{i = 1}^{n}{\left(1-\left(\frac{2}{3}\right)^{\av_{\pi}[c_{\pi}^{i-1}(\alpha)]}\right)}\right]}\label{EQ6}
\end{align}
Let $l(\alpha) \define \card{\set{\alpha' \in S: d(\alpha,\alpha') = 1}}$ denote that number of satisfying assignments of $\phi$ that has Hamming distance $1$ from $\alpha$. Using arguments from \cite{PPZ99} (cf. \cite{CIKP08}) we can bound $\av_{\pi}\left[d_{\pi}(\alpha)\right]$. For completeness we state it here. Given the definition of $l(\alpha)$, there are $n - l(\alpha)$ variables such that each of them appear as a unique true literal in some critical clause of $\phi$. It follows that each such variable $x_{\pi(i)}$ will be forced under randomly chosen $\pi \in S_n$ if $x_{\pi(i)}$ occurs last in the corresponding critical clause. This happens with probability at least $1/3$. Using linearity of expectation we have that expected number of forced variables is at least $((n - l(\alpha)))/3$, and hence,
\begin{align}
\av_{\pi}\left[d_{\pi}(\alpha)\right] \leq \left(n - \frac{(n - l(\alpha))}{3}\right)\label{EQ7}
\end{align}
Now we concentrate on giving bounds on $\av_{\pi}[c_{\pi}^{i-1}(\alpha)]$ for $i = 1, \ldots, n$. Let ${\mathcal C}(\alpha)$ be the set of all critical clauses of $\phi$ with respect to $\alpha$. Let us also denote by $r^{i}_{\pi}(\alpha)$ the number of critical clauses that are removed by $\PPZ$ at the end of $i$th execution of the for loop. Clearly the expected number of critical clauses in the beginning of the $i$th execution of the for loop, $\av_{\pi}[c_{\pi}^{i-1}(\alpha)]$ is equal to the expected number of critical clauses that were present in the beginning of the $(i-1)$th execution minus the expected number of critical clauses that were removed by $\PPZ$ at the end of the $(i-1)$th execution. It follows,$\av_{\pi}[c_{\pi}^{i-1}(\alpha)] = \av_{\pi}[c_{\pi}^{i-2}(\alpha)] - \av_{\pi}[r_{\pi}^{i-1}(\alpha)]$, with $\av_{\pi}[c_{\pi}^{0}(\alpha)] = c_{\pi}^{0}(\alpha)$. Let ${\mathcal C}^{i-2}_{\pi}(\alpha)$ denote the set of all critical clauses in the beginning of $i-1$th execution of the for loop. Also, let $X_c$ be an indicator random variable taking values in $\set{0,1}$ such that $X_c = 1$ iff clause $c \in {\mathcal C}^{i-2}_{\pi}(\alpha)$ is removed by the end of $i-1$th execution of the for loop. Using linearity of expectation we have that,
\begin{align}
\av_{\pi}[c_{\pi}^{i-1}(\alpha)] &= c_{\pi}^{i-2}(\alpha) - \sum\limits_{c \in {\mathcal C}^{i-2}_{\pi}(\alpha)}{\av_{\pi}[X_c]}\notag\\
&= c_{\pi}^{i-2}(\alpha) - \sum\limits_{c \in {\mathcal C}^{i-2}_{\pi}(\alpha)}{\left(1 \cdot \pr_{\pi}[X_c = 1] + 0 \cdot \pr_{\pi}[X_c = 0]\right)}\notag\\
&= c_{\pi}^{i-2}(\alpha) - \sum\limits_{c \in {\mathcal C}^{i-2}_{\pi}(\alpha)}{\pr_{\pi}[X_c = 1]}\notag
\end{align}
As discussed above, a clause can not be critical for more than one variable. On the other hand each variable $x_{\pi(i)}$ that appears as a unique true literal in some set of critical clauses of $\phi$ creates a partition of ${\mathcal C}(\alpha)$. Let us denote the cardinality of the partition of critical clauses corresponding to variable $x_{\pi(i)}$ with respect to $\alpha$ by $t^{i}_{\pi}(\alpha)$ (where, $t^{0}_{\pi}(\alpha) = 0$). Surely, $c_{\pi}^{0}(\alpha) = \sum_{i = 1}^{n}{t^{i}_{\pi}(\alpha)}$. Now in the $(i-1)$th execution we consider variable $x_{\pi(i-1)}$, that appears as a unique true literal in $t^{i-1}_{\pi}(\alpha)$ many critical clauses under assignment $\alpha$. There is one possible way a critical clause $c$ is removed by $\PPZ$ in accordance with assignment $\alpha$ under randomly chosen $\pi \in S_n$ - (as discussed above) when corresponding variable is forced, and probability of that event to occur for clause $c$ is at least $1/3$.\par
 Note that here we have ignored one particular effect of the statement $\phi := \phi[x_{\pi(i)} \leftarrow b]$. By this modification of $\phi$ in every execution of the for loop a critical clause with $3$ literals can become a clause having $2$ or less number of literals and still remain critical - but will not be considered in the deletion step in next execution of the for loop. However, considering this effect will only improve the success probability of return by $\DEL$, as there will be lesser number of critical clauses in the subsequent execution of the for loop, on the other hand it will make the analysis complicated.\par
\begin{figure}[htbp]
\centering
\includegraphics[viewport=0 0 400 300,width=0.6\textwidth,clip]{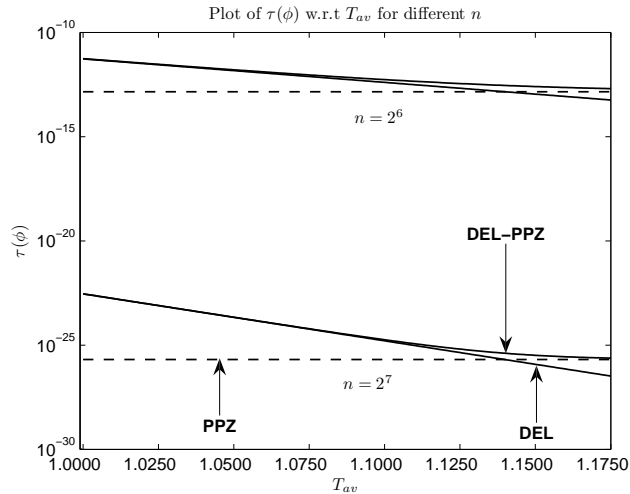}
\caption{Illustration of how success probability of $\PPZ$, $\DEL$ and $\DELPPZ$ changes with $1 \leq T_{av} < 2/(3 \cdot \log{(3/2)})$ for different values of $n$ with $s = 1$(Y-axis is in $\log$ scale).}
\label{FIG1}
\end{figure}

Based on the above discussion we have, $\sum_{c \in {\mathcal C}^{i-2}_{\pi}(\alpha)}{\pr_{\pi}[X_c = 1]} \geq {t_{\pi}^{i-1}(\alpha)}/{3}$. And we have, $\av_{\pi}[c_{\pi}^{0}(\alpha)] = c_{\pi}^{0}(\alpha),\text{ and } \av_{\pi}[c_{\pi}^{i-1}(\alpha)] \leq c_{\pi}^{i-2}(\alpha) - \frac{1}{3} \cdot t_{\pi}^{i-1}(\alpha)$. Solving this we obtain that,
\begin{align}
\av_{\pi}[c_{\pi}^{i-1}(\alpha)] \leq c_{\pi}^{0}(\alpha) - \frac{1}{3} \cdot \sum\limits_{j = 1}^{i-1}{t_{\pi}^{j}(\alpha)} &= c_{\pi}^{0}(\alpha) - \frac{1}{3} \cdot \sum\limits_{j = 1}^{n}{t_{\pi}^{j}(\alpha)} + \frac{1}{3} \cdot \sum\limits_{j = i}^{n}{t_{\pi}^{j}(\alpha)} \notag\\
= c_{\pi}^{0}(\alpha) - \frac{c_{\pi}^{0}(\alpha)}{3} + \frac{1}{3} \cdot \sum\limits_{j = i}^{n}{t_{\pi}^{j}(\alpha)} &= \frac{2}{3} \cdot c_{\pi}^{0}(\alpha) + \frac{1}{3} \cdot \sum\limits_{j = i}^{n}{t_{\pi}^{j}(\alpha)}\label{EQ8}
\end{align}
In following, we simplify notation by replacing $c_{\pi}^{0}(\alpha)$ with $c(\alpha)$, and $t_{\pi}^{i}(\alpha)$ with $t^{i}(\alpha)$. Now observe that in the expression $\prod_{j = 1}^{i-1}{(1-({2}/{3})^{\av_{\pi}[c_{\pi}^{j-1}(\alpha)]})}$ in Eq. (\ref{EQ6}), for every $i$, term $(1-({2}/{3})^{\av_{\pi}[c_{\pi}^{0}(\alpha)]})$ appears in every product. Also observe that $\sum_{j = i}^{n}{t_{\pi}^{j}(\alpha)} \geq 0$ for any $i$. So we use $\prod_{j = 1}^{i-1}{(1-({2}/{3})^{\frac{2}{3} \cdot c(\alpha)})} \leq \prod_{j = 1}^{i-1}{(1-({2}/{3})^{\av_{\pi}[c_{\pi}^{j-1}(\alpha)]})}$ as lower bound, and with Eq. (\ref{EQ7}) and Eq. (\ref{EQ8}) we modify Eq. (\ref{EQ6}) as follows:
\begin{align}
\tau(\phi) &\geq \sum\limits_{\alpha \in S}{\left[\left(\frac{2}{3}\right)^{c(\alpha)} \cdot \sum\limits_{i = 1}^{n}{\prod\limits_{j = 1}^{i-1}{\left(1-\left(\frac{2}{3}\right)^{\frac{2}{3} \cdot c(\alpha)}\right)}}\right]} +\notag\\
&\sum\limits_{\alpha \in S}{\left[2^{-\left(n - \frac{(n - l(\alpha))}{3}\right)} \cdot \prod\limits_{i = 1}^{n}{\left(1-\left(\frac{2}{3}\right)^{\frac{2}{3} \cdot c(\alpha)}\right)}\right]}\label{EQ9}
\end{align}
Now observe that since $\alpha$ is $(n - l(\alpha))$--isolated it must be that $c(\alpha) \geq (n - l(\alpha))$. In fact recall that $c(\alpha) = \sum_{i = 1}^{n}{t^{i}(\alpha)}$. Let us define $t$ as the minimum of $t^{i}(\alpha)$ over $i \in \set{1,\ldots,n}$ such that $x_{\pi(i)}$ appears as unique true literal in at least one critical clause, and $T_{av} \define c(\alpha)/(n - l(\alpha))$. We have $T_{av} (n - l(\alpha)) = c(\alpha) \geq t (n - l(\alpha))$. Also note that $t \geq 1$. Using these two facts with Eq. (\ref{EQ9}), we can lower bound $\tau(\phi)$ now as follows\footnote{All logarithms are base $2$.}:
\begin{align}
\tau(\phi) &\geq \sum\limits_{\alpha \in S}{\left[\left(\frac{2}{3}\right)^{T_{av} (n - l(\alpha))} \cdot \sum\limits_{i = 1}^{n}{\prod\limits_{j = 1}^{i-1}{\left(1-\left(\frac{2}{3}\right)^{\frac{2 t (n - l(\alpha))}{3}}\right)}}\right]} + \notag\\
&\sum\limits_{\alpha \in S}{\left[2^{-\left(n - \frac{(n - l(\alpha))}{3}\right)} \cdot \prod\limits_{i = 1}^{n}{\left(1-\left(\frac{2}{3}\right)^{\frac{2 t (n - l(\alpha))}{3}}\right)}\right]}\notag\\
&= 2^{-T_{av} \cdot n \cdot \log{(3/2)}} \cdot \sum\limits_{\alpha \in S}{\left[\left(\frac{2}{3}\right)^{-T_{av} \cdot l(\alpha)} \cdot \left(1-\left(\frac{2}{3}\right)^{\frac{2 t (n - l(\alpha))}{3}}\right)^{n-1}\right]} + \notag\\
& 2^{-\frac{2n}{3}} \cdot \sum\limits_{\alpha \in S}{\left[2^{-\frac{l(\alpha)}{3}} \cdot \left(1-\left(\frac{2}{3}\right)^{\frac{2 t (n - l(\alpha))}{3}}\right)^{n}\right]}\label{EQ10}
\end{align}
Let $L \define \av_{\alpha \in S}[l(\alpha)]$ and $s \define \card{S}$. Using Jensen's inequality we obtain:
\begin{align}
&\sum\limits_{\alpha \in S}{\left[\left(\frac{2}{3}\right)^{- T_{av} \cdot l(\alpha)} \cdot \left(1-\left(\frac{2}{3}\right)^{\frac{2t(n - l(\alpha))}{3}}\right)^{n-1}\right]} \geq\notag\\
&s \cdot \left(\frac{2}{3}\right)^{- T_{av} \cdot L} \cdot \left(1-\left(\frac{2}{3}\right)^{\frac{2t(n - L)}{3}}\right)^{n-1} \label{EQ11}
\end{align}
And,
\begin{align}
\sum\limits_{\alpha \in S}{\left[2^{-\frac{l(\alpha)}{3}} \cdot \left(1-\left(\frac{2}{3}\right)^{\frac{2t(n - l(\alpha))}{3}}\right)^n \right]} \geq s \cdot 2^{-L/3} \cdot \left(1-\left(\frac{2}{3}\right)^{\frac{2t(n - L)}{3}}\right)^n \label{EQ12}
\end{align}
Combining Eq. (\ref{EQ11}) and Eq. (\ref{EQ12}) with Eq. (\ref{EQ10}) we have:
\begin{align}
\tau(\phi) &\geq s \cdot \left(\frac{2^{-\left(n-L\right) \cdot T_{av} \cdot \log{(3/2)}}}{\left(1-\left(\frac{2}{3}\right)^{\frac{2t(n - L)}{3}}\right)} + 2^{-\left(2n + L\right)/3}\right) \cdot \left(1-\left(\frac{2}{3}\right)^{\frac{2t(n - L)}{3}}\right)^n \notag\\
&\geq s \cdot \left(2^{-\left(n-L\right) \cdot T_{av} \cdot \log{(3/2)} + o(1)} + 2^{-\left(2n + L\right)/3}\right) \cdot \left(1-\left(\frac{2}{3}\right)^{\frac{2t(n - L)}{3}}\right)^n \label{EQ13}
\end{align}
In order to bound $L$ we will use the edge isoperimetric inequality from \cite{Har67}, which states that for any $S \subseteq \set{0,1}^n$, $\card{\set{(a,a') | a, a' \in S \text{ and } d(a,a') = 1}} \leq \card{S} \cdot \log{(\card{S})}$, and $\sum_{\alpha \in S}{l(\alpha)} \leq s \cdot \log{s}$. So using this result as in \cite{CIKP08} $L = \av_{\alpha \in S}[l(\alpha)] \leq \log{s}$. On the other hand, it is not hard to observe that the lower bound on $\sum_{\alpha \in S}{l(\alpha)}$ is $0$ as long as $s \leq 2^{n-1}$. This can be seen as follows. We consider $\set{0,1}^n$ as the vertex set of a graph (Hamming cube, denoted $Q_n$) and for $a, a' \in \set{0,1}^n$, $aa'$ is an edge of this graph iff $d(a,a') = 1$. Now the lower bound in question corresponds to finding a subgraph of $Q_n$ having $s$ many vertices and having minimum number of induced edges having both of their end-points in $S \subseteq \set{0,1}^n$. Now observe that since $Q_n$ is bipartite, with $s \leq 2^{n-1}$ we have always a set of vertices of size $s$ having no edges between them. Updating Eq. (\ref{EQ13}) with this we have:
\begin{align}
\tau(\phi) &\geq s \cdot \left(2^{-n \cdot T_{av} \cdot \log{(3/2)}} + 2^{-\left(2n + \log{s}\right)/3}\right) \cdot \left(1-\left(\frac{2}{3}\right)^{\frac{2t(n - \log{s})}{3}}\right)^n \notag\\
&= \left(s \cdot 2^{-n \cdot T_{av} \cdot \log{(3/2)}} + \left(2^{-n} \cdot s\right)^{2/3}\right) \cdot \left(1-\left(2^{-n} \cdot s\right)^{2/3 \cdot t \cdot \log{(3/2)}}\right)^n\label{EQ14}
\end{align}
Now it can be seen that the term,$(1-(2^{-n} \cdot s)^{2/3 \cdot t \cdot \log{(3/2)}})^n$ converges to $1$ very fast with $n$. So for sufficiently large $n$ we can ignore this term. Thus for sufficiently large $n$ we have from Eq. (\ref{EQ14}), 
\begin{align}
\tau(\phi) \geq \left(s \cdot 2^{-n \cdot T_{av} \cdot \log{(3/2)}} + \left(2^{-n} \cdot s\right)^{2/3}\right)\label{EQ15}
\end{align}
Lower bound on $\tau(\phi)$ from Eq. (\ref{EQ15}) shows that (like $\PPZ$ \cite{PPZ99}) performance of the algorithm $\DELPPZ$ improves with more number of solutions. On the other hand for any value of $1 \leq T_{av} < 2/(3 \cdot \log{(3/2)}) = 1.13967$, performance of $\DELPPZ$ is better than $\PPZ$.  For higher values of $T_{av}$ and with $s = 1$ performance of the algorithm $\DELPPZ$ tends to become same as the performance of $\PPZ$ algorithm, which is $1.5875^{-n}$. On the other hand for $s = 1$ (unique solution) and $T_{av} = 1$ (one critical clause per variable) performance of the algorithm $\DELPPZ$ tends to become same as the performance of algorithm $\DEL$, which is $1.5^{-n}$ (see Fig. \ref{FIG1}.).
Our results on the algorithm $\DELPPZ$ can now be summarized in the following statements:
\begin{lemma}
Let $\phi$ be any $3$-$\CNF$ formula over $n$ variables that has $s$ number of satisfying assignments, and let $T_{av}$ be the average number of clauses per variable that appear as unique true literal in one or more critical clauses in $\phi$. Then probability that one iteration of algorithm $\DELPPZ$ outputs some satisfying assignment is at least,
\begin{align}
\left(s \cdot 2^{-n \cdot T_{av} \cdot \log{(3/2)}} + \left(2^{-n} \cdot s\right)^{2/3}\right)\notag
\end{align}
\end{lemma}
\begin{theorem}
Let $\phi$ be any $3$-$\CNF$ formula over $n$ variables and let $T_{av} \in [1,2/(3 \cdot \log{(3/2)})]$ be the average number of clauses per variable that appear as unique true literal in one or more critical clauses in $\phi$. Then probability that one iteration algorithm $\DELPPZ$ outputs some satisfying assignment is at least $1.5^{-n}$ for $T_{av} = 1$ and decreases to $1.5875^{-n}$ for $T_{av} = 2/(3 \cdot \log{(3/2)})$. For $T_{av} > 2/(3 \cdot \log{(3/2)})$ probability that one iteration algorithm $\DELPPZ$ outputs some satisfying assignment is at least $1.5875^{-n}$. And these bounds are tight for $\phi = \wedge_{i =1}^{m-1}{(x_{3i} \oplus x_{3i+1} \oplus x_{3i+2})}$ where $n = 3m$.
\end{theorem}
Now recall that we can also bound the error probability of the algorithm to $o(1)$ if we execute the algorithm $\DELPPZ$ for $\omega \geq n/\tau(\phi)$ times. With this we obtain following results:
\begin{theorem}
Let $T_{av} \geq 1$ be a real number. There is a randomized algorithm for $3$-$\SAT$, namely $\DELPPZ$, that given any $3$-$\CNF$ formula $\phi$ over $n$ variables with $s$ number of satisfying assignments, makes one sided error of at most $o(1)$ on satisfiable instances, otherwise outputs one of the satisfying assignments of $\phi$ in expected time 
\begin{align}
{\mathcal O}\left(\min{\set{\left(\poly(n) \cdot \left(\frac{2^{n \cdot T_{av} \cdot \log{(3/2)}}}{s}\right)\right),\left(\poly(n) \cdot \left(\frac{2^{n}}{s}\right)^{2/3}\right)}}\right)\notag
\end{align}
\end{theorem}
\section{Concluding remarks}
\label{SEC4}
As stated in the introduction that recently best known randomized bound for $3$--$\SAT$ is ${\mathcal O}\left(1.32216^n\right)$ \cite{Rolf06}. It is interesting to note that this algorithm is a combination of the random walk algorithm of \cite{Uwe99,Sch02} and algorithm of \cite{PPSZ98} (we will call this algorithm $\PPSZ$), and success probability of algorithm in \cite{Rolf06} is maximum of the success probability of random walk algorithm of \cite{Uwe99,Sch02} and algorithm $\PPSZ$. Algorithm $\PPSZ$ is a combination of $3^d$ bounded resolution on input $3$-$\CNF$ formula $\phi$ followed by the $\PPZ$ algorithm. Purpose of using a bounded resolution first is to increase the success probability of $\PPZ$ algorithm - by increasing the number of critical clauses per variable - as that will in effect increase the probability that a variable (that appears as unique true literal in a set of critical clauses) is forced with respect to a randomly chosen permutation. On the other hand algorithm $\DELPPZ$ performs better when the average number of critical clause per variable in $\phi$ is close to $1$. We believe that for values of $T_{av}$ close to $1$ our algorithm improves the algorithm $\PPSZ$ and best known randomized bound for $3$--$\SAT$ as presented in \cite{Rolf06}. We will consider this analysis as our future work.
%
%
%
\bibliographystyle{splncs}
\bibliography{sat}  


\end{document}